\documentclass[graybox]{svmult}

\usepackage{type1cm}        
\usepackage{makeidx}         
\usepackage{graphicx}        
\usepackage{multicol}        
\usepackage[bottom]{footmisc}

\usepackage{newtxtext}       %
\usepackage[varvw]{newtxmath}       

\makeindex             

\begin{document}

\title*{Memories of Gravitational Shockwaves}
\author{Dmitri V. Fursaev\orcidID{0000-0002-5693-3112}}
\institute{D.V. Fursaev \at Bogoliubov Laboratory of Theoretical Physics, Joint Institute for Nuclear Research, 141 980, Dubna, Moscow Region, Russian Federation, \email{fursaev@theor.jinr.ru}}
%
%
\maketitle

\abstract*{.}

\abstract{Gravitational shockwaves produce perturbations of field systems. We study classical scalar and electromagnetic fields and  gravitational memory effects left after the action on the fields of plane-fronted gravitational shockwaves. The gravitational memory plays a key role for the choice of Cauchy data which determine the perturbations.
We demonstrate that field systems`remember' only a spatial `profile' of the shock, but not the form of the signal. Moreover the dependence 
on the spatial profile  can be expressed in a geometric way as a transformation of fields under coordinate supertranslations defined near the shockwave front. We also discuss applications of our results to astrophysics. }

\section{Introduction}
\label{sec:1}

Gravitational shockwaves (GSW) are solutions to the Einstein equations which describe short gravitational pulses with rapid oscillations of the Riemann curvature on the shockwave fronts. Inside the shockwave the gravity is strong, as opposed to weak gravitational waves, which are
ripples of the space-time ruled by the linearized  Einstein equations.  

Gravitational shockwaves produce perturbations of classical electromagnetic (EM), scalar, or gravitational fields, which may result in interesting physical effects.  Here we study perturbations generated by plane-fronted shockwaves which propagate in flat   space-times. We suppose that gravitational shockwaves are created by different sources which move with the speed of light. 
One may expect that this situation model gravitational waves produced  
by the motion of a broader class of ultra-relativistic objects. The geometries we consider here have a unique feature: their Einstein tensor is linear in the metric.  

Suppose a plane-fronted shockwave is of a `sandwich' type, that is, it is concentrated between the two fronts (null hypersurfaces): the past, $u=\delta$, and the future, $u=0$, fronts, where $u$ is a retarded time and $\delta>0$.
Consider a free field $\phi$ (which is a scalar or EM field). The problem for perturbations  $\hat{\phi}$ of $\phi$ caused by the GSW is a the following characteristic Cauchy problem in the causal past of the GSW:
\begin{equation}\label{i.1}
P~\hat{\phi}=0~~,~~u\geq \delta~~,
\end{equation}
\begin{equation}\label{i.2}
\hat{\phi}=\phi_0~~,~~u=\delta~~,
\end{equation}
where $P$ is a corresponding relativistic second order hyperbolic differential operator. The problem can be solved
once the Cauchy data $\phi_0$ are known. Suppose $\bar{\phi}$ is the unperturbed field just before the shock, at $u= 0$.
To solve (\ref{i.1}), (\ref{i.2}) 
one needs to determine how $\bar{\phi}$ transforms under the action of the GSW in the interval  $0< u <\delta$. 
In general, this may be a difficult task since the gravity in this interval 
is strong and field equations are essentially nonlinear in the metric.  

To put it another way, we pose the question: how does $\phi_0$ `remember' shock gravitational waves?
We show that for a class of GSW described above field perturbations do not `remember' the shape of the gravitational signal but depend on its spatial profile.
Moreover, the gravitational memory effect appears as a special coordinate transformation near the past GSW front:
\begin{equation}\label{i.3}
\phi_0={\cal L}_\zeta\bar{\phi}~~,~~u=\delta~~.
\end{equation}
The operation ${\cal L}_\zeta$ is the Lie derivative along a vector field $\zeta$, and on the shockwave front $\zeta$ coincides with generator of well known Penrose supertranslations \cite{Penrose:1972xrn}.

The rest of the work is organized as follows.  In Section \ref{sec:2} we give necessary definitions and describe properties of plane-fronted shockwaves, including impulsive limit $\delta \to 0$, when the GSW geometry is treated in a sense of distributions. 
We start, in Section \ref{sec:3} with the memory effect for trajectories of test particles under the action of the GSW. In our approximation, test particles before and after the shockwave pulse move along geodesics in the Minkowsky space-time.
After the shockwave, mutual orientations of 4-velocities $\bar{u}^\mu$ of the particles change  as
\begin{equation}\label{i.4}
\delta u^\mu={\cal L}_\zeta \bar{u}^\mu~~,~~u=\delta~~.
\end{equation}
We then
continue with perturbations of scalar fields and show that the effect of GSW is completely determined by  the  supertranslations identical to those which appear in the impulsive limit.  Perturbations of EM fields by the shockwaves are analyzed 
in Section \ref{sec:4}. Due to the gauge invariance solutions to the Maxwell equations can be uniquely determined by components of the Maxwell strength which are normal to constant $u$ null hypersurfaces. We demonstrate that the Cauchy problem for these components
has the form of (\ref{i.1})-(\ref{i.3}) . 
Concluding remarks with a review of possible applications of these results are presented in Section \ref{sec:5}.

\section{Plane-fronted gravitational shockwaves}
\label{sec:2}

We start with necessary definitions. For a gravitational shockwave moving along 
the $x$ direction, from the left to the right, the space-time metric is chosen to be
\begin{equation}\label{1.1}
ds^2=-dv du -H(v,u,y)du^2+d y_i^2~.
\end{equation}
The retarded  and advanced time coordinates are  $u=t-x$, $v=t+x$. With vectors $l_\mu=-\delta^u_\mu$, which are null normals to constant $u$ hypersurfaces, the metric tensor can be written as $g_{\mu\nu}=\eta_{\mu\nu}-Hl_\mu l_\nu$, where $\eta_{\mu\nu}$ is the Minkowsky metric.
The properties of the wave are described by $H$ which we take as 
\begin{equation}\label{1.1a}
H(v,u,y)=\chi(u) f(v,y)~~,
\end{equation}
where $f(v,y)$ is called the 'profile' function. Evolution of the shock in the retarded time is determined by 
$\chi(u)$ which can be called the 'signal' function.  We also define the integrated signal function
\begin{equation}\label{1.3}
\bar{\chi}(u)=\int^u_{-\infty} \chi(u')du'~.
\end{equation}
To normalize the profile and signal functions we assume that $\bar{\chi}(u)\to 1$ at $u\to \infty$.
In most important cases listed below (which are a subclass of Kundt space-times \cite{Kundt:1961}) $f$ does not depend on $v$. Thus, one can call $\chi(u)$ and $f$ the time and spatial profiles of the GSW, respectively.

\begin{table}[!t]
\caption{Examples of ultra-relativistic  sources with energy distribution $\sigma$ . The parameters $E_p$, $E_m$, $E_s$, $E_w$ are energies of the sources, $\rho=(y_i^2)^{1/2}$. }
\label{tab:1}       
%
%
\begin{tabular}{p{4cm}p{2cm}p{2cm}p{2cm}}
\hline\noalign{\smallskip}
Ultrarelativistic  source & Trajectory  & $\sigma(y)$  & $f(y)$ \\
\noalign{\smallskip}\svhline\noalign{\smallskip}
Massless point particle   &  $u=\rho=0$ &  $E_p\delta(\rho)$  & $8 G E_{p} \ln \rho $ \\
Global monopole  &  $u=\rho=0$  & $E_{m}/\rho$ & $16\pi G E_{m}\rho $ \\
Null straight cosmic string  &  $u=y_1=0$ & $E_{s}\delta(y_1)$ & $ 8\pi G E_s |y_1| $ \\
Domain wall & $u=0$ & $E_{w}$ & $4\pi G E_{w} \rho^2 $  \\
\noalign{\smallskip}\hline\noalign{\smallskip}
\end{tabular}
\end{table}

In this paper we consider gravitational sandwich-like perturbations whose signal function is non-vanishing only in some time interval, say, when $0<u<\delta$. Inside interval $\chi(u)$ can be absolutely arbitrary.
At $u<0$, $u>\delta$ metric \eqref{1.1} coincides with the Minkowsky metric. As has been mentioned above, the important property of 
\eqref{1.1}  is that non-vanishing components of its Einstein tensor are linear in $H$, 
\begin{equation}\label{1.6}
	G_{uu}=\frac 12 \partial^2_{i} H~,~
	G_{ui}=\partial_v\partial_i H~,~
	G_{ij}= 2\delta_{ij}\partial_v^2 H~.
\end{equation}
Therefore in the Einstein theory
the shockwaves are produced by sources with the stress-energy tensor
\begin{equation}\label{1.7}
	T_{\mu\nu}=\chi(u)\left (\sigma l_\mu l_\nu+j_i ( l_\mu e_\nu^i+ l_\nu e_\mu^i)+p\delta_{ij}e_\mu^ie_\nu^j\right)~,
\end{equation}
$$
\sigma=f_{,ii}/\kappa~~,~~j_i=-2f_{,vi}/\kappa~~,~~p=4f_{,vv}/\kappa~~,~~\kappa=16\pi G~~.
$$
Vectors $e^i_\mu=\delta^i_\mu$ are tangent to constant $u$ null hypersurfaces.  Examples of the profile functions
for different sources are given in Table \ref{tab:1}.

Note that in addition to our choice of $H$ in (\ref{1.1a}) there are other types of sandwich-like gravitational plane  waves. For example,
one can take $H=h_{ij}(u)y^iy^j$, see \cite{Gibbons:1975jb},\cite{Garriga:1990dp}. The source of such waves is a sequence of domain  walls with the surface energies $\sigma=\frac 12 h_{ii}(u)$. We do not consider these metrics here.

An immediate consequence of \eqref{1.6} is that
 $G_{\mu\nu}$ is well-defined in the limit of extremely short signals, when $\chi(u)\to \delta(u)$.  The wave front of GSW is then localized exactly on the null hypersurface $u=0$.
 
A singular geometry sourced by (\ref{1.7}) with $\chi(u)=\delta(u)$ can be described by using an approach by Penrose  \cite{Penrose:1972xrn}
and soldering transformations of \cite{Blau:2015nee}. The idea is as follows.
One cuts $R^{1,3}$ by the front $u=0$ into two 'halfs', ${\cal M}^+$ ($u>0$) and
${\cal M}^-$ ($u<0$). Then ${\cal M}^\pm$ are ``glued'' again along $u=0$ with the following identification of points:
\begin{equation}\label{1a.9}
v_+= v-f(v,y)~~,~~y^i_+=y^i~~.
\end{equation}
Supertranslations (\ref{1a.9}) leave invariant metric on $u=0$ and, thus, make an infinite
dimensional group of isometries of $u=0$, also known as Carroll transformations
\cite{Duval:2014lpa,Duval:2014uoa}.

With the help of  a null vector $n$ at $u=0$, such as $(n\cdot l)=-1$,
$(n\cdot e_i)=0$, one defines the transverse curvature, introduced in \cite{Barrabes:1991ng},\cite{Poisson:2002nv},
\begin{equation}\label{1a.10}
{\cal C}_{ab}=n_{\mu;\nu} e_a^\mu  e_b^\nu~~,
\end{equation}
where $e_v=l, e_i=e^i$.  Then
\begin{equation}\label{1a.11}
\sigma={2 \over \kappa}[{\cal C}_{ii}]~~,~~j_i=-{2 \over \kappa}[{\cal C}_{vi}]~~,~~p={2 \over \kappa}[{\cal C}_{vv}]~~,
\end{equation}
\begin{equation}\label{1a.12}
[{\cal C}_{ab}]\equiv {\cal C}_{ab}^+-{\cal C}_{ab}^-~~,
\end{equation}
where ${\cal C}_{ab}^\pm$ correspond to ${\cal M}^\pm$.  In the considered case
\begin{equation}
    [{\cal C}_{vv}]=2[\partial_u g_{vv}]~~,~~
    [{\cal C}_{vi}]=[\partial_u g_{vi}]~~,~~
    [{\cal C}_{ii}]=\frac{1}{2}[\partial_u g_{ii}]~~.
\end{equation}
To know $\partial_u g_{\mu\nu}^{~+}$
on $u=0$ one needs coordinate transformations of ${\cal M}^+$
\begin{equation}\label{1a.13}
x^\mu_+=x^\mu-\zeta^\mu(x)~~,
\end{equation}
\begin{equation}\label{1.14}
\zeta^\mu(x)\simeq f\delta^\mu_v+\frac u2 \eta^{\mu a}f_{,a}~~.
\end{equation}
which are an ``extension'' of the supertranslations to a neighborhood of $u=0$.
Vector  $\zeta^\mu(x)$ is completely determined \cite{Blau:2015nee} by the requirement that
(\ref{1.14}) do not change the metric $g_{\mu\nu}$ at $u=0$.
Since in a vicinity of $u=0$ deviation of the metric from the flat one is nontrivial,
\begin{equation}\label{1a.15}
g_{\mu\nu}^+ \simeq g_{\mu\nu}^-+{\cal L}_\zeta g_{\mu\nu}^{~-}=
\eta_{\mu\nu}+\zeta_{\mu,\nu}+\zeta_{\nu,\mu}~~,
\end{equation}
one finds:
\begin{equation}\label{1a.16}
\partial_u g_{v v}^{~+}|_{u=0}=f_{,vv}~~,~~
\partial_u g_{v i}^{~+}|_{u=0}=f_{,vi}~~,~~
\partial_u g_{ij}^{~+}|_{u=0}=f_{,ij}~~.
\end{equation}
Substitution of (\ref{1a.16}) in  (\ref{1a.11}) yields (\ref{1.7}).

\section{Test particles and scalar fields after gravitational shocks}
\label{sec:3}

Consider a particle with 4-velocity $\bar{u}^\mu$ before it meets the shockwave. 
According to geodesic equations the velocity changes due to interaction with the shock.
In the linear in $H$ approximation
\begin{equation}\label{1.2}
{d u^\mu \over du}=- l^\mu (H_{,\nu}\bar{u}^\nu)-\frac 12 H^{,\mu}
+O(H^2)~,
\end{equation}
where we took into account that $(\bar{u}\cdot l)=-1$. It follows from \eqref{1.3}, \eqref{1.2} that the velocity acquires 
two parts 
$$
u^\mu\simeq \bar{u}^\mu+\chi(u) u^\mu_1+\bar{\chi}(u)u^\mu_2~~,
$$
where
\begin{equation}\label{1.4}
u^\mu_1=-\frac 12 l^\mu f~,~u^\mu_2=-\frac 12 (f_{,\nu}\bar{u}^\nu)-\frac 12 l^\mu f~.
\end{equation}
The term $\chi(u) u^\mu_1$ in the velocity implies that the advanced time on the particle trajectory changes as 
$\delta v(u)=-\bar{\chi}(u)f$, in agreement with (\ref{1a.9}) .  Terms proportional to $\chi(u)$ disappear after a time  interval exceeding $\delta$.

We suppose that the profile function changes slowly during the action of the shock, $ |f_{,\mu}|\delta \ll |f|$.  Then residual transformations of the particle coordinates and the velocity just after the shock can be written as a pure coordinate transformations 
(\ref{i.4}) generated by vector (\ref{1.14}).
To see this one should take into account that $\bar{\chi}(u)=1$ at $u>\delta$.

Thus, the gravitational memory effect for particles moving through the shockwave has a pure geometrical form and is  determined by the profile of the wave $f$.  Particles do not `remember' the form of the gravitational signal $\chi(u)$.  

For different type of  impulsive plane  waves, waves given by (\ref{1.1}) with $H=h_{ij}(u)y^iy^j$, the gravitational memory effect for particles has been 
studied in \cite{Zhang:2017jma},\cite{Steinbauer:2018iis}.

Consider now a free scalar field $\phi$ with the equation:
\begin{equation}\label{2.1}
(-\Box+m^2)\phi=j~~,
\end{equation}
where $j$ is a source. If one takes into into account that 
$g^{\mu\nu}=\eta^{\mu\nu}+H l^\mu l^\nu$, $\det g_{\mu\nu}=-1/4$ and $l^\mu \partial_\mu=2\partial_v$ the operator in (\ref{2.1}) can be written as
$$
\Box=\partial_\mu g^{\mu\nu} \partial_\nu=-\bar{\Box}+4 \partial_v H  \partial_v~~.
$$
$\bar{\Box}=-4\partial_u\partial_v+\partial_i^2$ is defined for the flat metric. Suppose $\bar{\phi}$ is a solution in the Minkowsky spacetime
\begin{equation}\label{2.2}
(-\bar{\Box}+m^2)\bar{\phi}=\bar{j}~~.
\end{equation}
We define perturbation due to the shockwave as a difference 
\begin{equation}\label{2.3}
\hat{\phi}\equiv \phi-\bar{\phi}~~,~~u>\delta~~,
\end{equation}
by assuming that 
\begin{equation}\label{2.4}
j=\bar{j}~~,~~u>\delta~~.
\end{equation}
The equation for the perturbations, which follows from (\ref{2.1})-(\ref{2.3}), is
\begin{equation}\label{2.5}
(-\bar{\Box}+m^2)\hat{\phi}=J+\chi(u)J_g(\phi,f)~~,
\end{equation}
\begin{equation}\label{2.6}
J=j-\bar{j}~~,~~J_g(\phi,f)=- \partial_v(f  \partial_v \phi)~~.
\end{equation}
Here $J$ takes into account an action of the GSW on the current. Since we are interested in the linear in $f$ effects one can approximate $J_g(\phi,f)\simeq J_g(\bar{\phi},f)$.
Then the term $\chi(u)J_g(\bar{\phi},f)$ is an effective current induced by rapidly changing gravitational field on steady scalar background $\bar{\phi}$. 

Now one can proceed as in the case of geodesics and look for a solution to (\ref{2.5}) during the action of the shock in the form
\begin{equation}\label{2.7}
\hat{\phi} \simeq \chi(u) \phi_1+\bar{\chi}(u)  \phi_2~~,~~0 < u < \delta~~,
\end{equation}
where dependence on $u$ of functions $\phi_k$ can be neglected. The same reasonigs also require that in this interval
\begin{equation}\label{2.8}
J \simeq \chi(u) i_1+\bar{\chi}(u)  i_2~~.
\end{equation}
However $i_2=0$, according to our definition of the perturbation, see (\ref{2.4}).

If (\ref{2.7}), (\ref{2.8}) are substituted into (\ref{2.5}) there appear terms, in the right hand and left hand sides of the equation, 
which are
proportional to $\partial_u \chi(u), \chi(u), \bar{\chi}(u)$. Hence one comes to constraints:
\begin{equation}\label{2.9}
\partial_v \phi_1=0~~,
\end{equation}
\begin{equation}\label{2.10}
4\partial_v \phi_2-(\partial_i-m^2)\phi_1=i_1+4 \partial_v(f  \partial_v \bar{\phi})~~,
\end{equation}
which do not depend on the time profile of the GSW. Solution to (\ref{2.9}) is trivial, $\phi_1=0$, since perturbation is zero in the far past,
as $v\to -\infty$. Then $4\partial_v \phi_2=i_1+4 \partial_v {\cal L}_\zeta \bar{\phi}$ and just behind the wave front 
\begin{equation}\label{2.11}
\hat{\phi}={\cal L}_\zeta \bar{\phi}+\phi_c~~,~~u\simeq \delta~~,
\end{equation}
where $\phi_c$ is related to the change of the source on the wavefront 
\begin{equation}\label{2.12}
\phi_c(v,y)=\frac 14 \int_{-\infty}^vdv' i_1(v',y)~~.
\end{equation}
One can call  $\phi_c$ a contact term since it comes out from points where the source of the field contacts the wave front.
The contact terms are not related to the gravitational memory of the field. If these terms are omitted, (\ref{2.11}) coincides with (\ref{i.2}),(\ref{i.3}). For a point-like source of the field terms $J$ and $\phi_c$ are related to a change of the trajectory of the source. 

Effects of gravitational shockwaves on scalar fields have been studied in many publications starting with \cite{tHooft:1985NPB}.
Much attention to this effect has been attracted in the context of the graviton dominance in scattering of ultra-high energy 
particles \cite{tHooft:1987vrq}. Scalar fields were considered there as plane waves $\bar{\phi}=e^{ikx}$, while  shockwaves were taken in the impulsive limit. In this case condition (\ref{2.11}) (without contact terms) implies transformation of particle's 4-momentum $k^\mu$ in
the complete agreement with results of this Section. A comprehensive analysis of scalar field perturbations generated  by gravitational shockwaves in the impulsive limit along with a discussion of how contact terms are taken into account in the Cauchy data can be found in \cite{Fursaev:2024czx}.

\section{Gravitational shocks and Maxwell fields}
\label{sec:4}

Now we extend this analysis to the case of EM fields under the action of the GSW.
Consider Maxwell equations on shockwave spacetime \eqref{1.1}  and in the Minkowsky spacetime
\begin{equation}\label{1.15}
\partial_\mu F^{\mu\nu}=j^\nu~,~\partial_\mu \bar{F}^{\mu\nu}=\bar{j}^\nu ~,
\end{equation}
respectively.  As a result of the specific structure of \eqref{1.1} the both equations look similar, 
but indexes of $F^{\mu\nu}$ are risen with the help of the inverse metric tensor $g^{\mu\nu}$.
Also conservation laws for the currents are: 
$$
\partial_\mu j^\mu=\partial_\mu \bar{j}^\mu=0~~.
$$

Let $\bar{F}^{\mu\nu}$ be the Maxwell strength of EM field before the shock. We define its perturbation as the difference  
$$
\hat{F}_{\mu\nu}\equiv F_{\mu\nu}-\bar{F}_{\mu\nu}~~,~~u>\delta~~.
$$
One can show that \eqref{1.15} lead to equations 
\begin{equation}\label{1.16}
\partial_\mu \hat{F}^{\mu\nu}=J^\nu+\chi(u) J^\nu_g(F,f)~,
\end{equation}
where $J^\nu=j^\nu-\bar{j}^\nu$ is a variation of the current during the shock, and
\begin{equation}\label{1.17}
J^\nu_g(F,f)=\partial_\mu\left[f l^\alpha (l^\nu \eta^{\mu\beta}-l^\mu \eta^{\nu\beta}) F_{\alpha\beta})\right]~.
\end{equation}
It is clear that $\partial_\mu J^\mu=\partial_\mu (\chi(u) J^\mu_g)=\partial_\mu J^\mu_g=0$ and $l_\mu J^\mu_g=0$. In the first approximation one can put
$J^\nu_g(F,f)\simeq J^\nu_g(\bar{F},f)$. After some algebra one can write 
\begin{equation}\label{1.17a}
J^\nu_g(\bar{F},f)=- 2f l_\mu {\cal L}_\zeta \bar{F}^{\mu\nu}-fl^\nu l_\alpha \partial_\lambda \bar{F}^{\alpha\lambda}~.
\end{equation}
One can interpret $\chi(u) J^\nu_g(\bar{F},f)$ as an effective current, tangent to constant $u$ null hypersurfaces in the region 
$0 < u < \delta$ of the GSW pulse. 
This current is generated by rapidly changing gravitational field on a smooth EM background $\bar{F}_{\mu\nu}$. 
The mechanism which yields $\hat{F}^{\mu\nu}$ is analogous to the well-known generation of transition radiation \cite{Ginzburg:1945zz},\cite{Ginzburg:1979wi} when a charged particle crosses the boundary of two physical media. Therefore one can also call $\hat{F}^{\mu\nu}$ transition radiation generated by gravitational shockwaves. 

Our aim is to show that the transition radiation 
on shockwaves is determined by the gravitational memory effect and is expressed in a form of coordinate transformation \eqref{i.3}.
We suppose that the transition radiation and variation of the currents in the interval $0\leq u \leq \delta$ have the form:
\begin{align}
	\hat{F}^{\mu\nu}&\simeq \chi(u) p^{\mu\nu}+\bar{\chi}(u) q^{\mu\nu}~, \label{1.18}
	\\
	J^{\mu}&\simeq \chi(u) i_1^{\mu}+\bar{\chi}(u) i_2^{\mu}~, \label{1.19}
\end{align}
similar to variations of 4-velocities of particles. Here $p^{\mu\nu}$, $q^{\mu\nu}$, $i_k^{\mu}$ are some quantities which change slowly on the above interval.  The gravitational memory effect is determined by $q^{\mu\nu}$ and $i_2^{\mu}$.

By requiring that terms at $\partial_u\chi(u)$, $\chi(u)$, $\bar{\chi}(u)$ in the Bianchi identities and in the conservation law $\partial_\mu J^\mu=0$  vanish one gets the restrictions:
\begin{gather}
	p_{(\mu\nu}l_{\rho)}=0~,~p_{(\mu\nu,\rho)}=q_{(\mu\nu}l_{\rho)}~, \label{1.20}
	\\ 
	l_\mu i_1^{\mu}=0~,~\partial_\mu i_1^{\mu}=l_\nu i_2^\nu~,~\partial_\mu i_2^{\mu}=0~, \label{1.21}
\end{gather}
where symbol $(\mu\nu...\rho)$ denotes complete symmetrization of the indices. Indices are risen and lowered with the flat metric.
Substitution of \eqref{1.18}, \eqref{1.19} to Maxwell equations \eqref{1.16} yields:
\begin{align}
	l_\mu p^{\mu\nu}&=0~, \label{1.22}
	\\
	\partial_\mu p^{\mu\nu}-l_\mu q^{\mu\nu}&= i_1^{\nu}+J_g^\nu~, \label{1.23}
	\\
	\partial_\mu q^{\mu\nu}&=i_2^\nu~. \label{1.24}
\end{align}
From \eqref{1.20}, \eqref{1.22} we find
\begin{gather}
	p_{\mu\nu}=\sigma_i(l_\mu e^i_\nu-l_\nu e^i_\mu)~, \label{1.25}
	\\
	q_{\mu\nu}=e^i_\mu \sigma_{i,\nu}- e^i_\nu\sigma_{i,\mu}+\rho_i(l_\mu e^i_\nu-l_\nu e^i_\mu)~, \label{1.26}
\end{gather}
where $\sigma_i$ and $\rho_i$ are some functions. Equations  \eqref{1.25}, \eqref{1.26}  imply that
\begin{equation}\label{1.27}
\partial_\mu p^{\mu\nu}+l_\mu q^{\mu\nu}=-\sigma_{i,i}l^\nu~,
\end{equation}
which together with \eqref{1.17a}, \eqref{1.25} yields
\begin{gather}
	l_\mu q^{\mu\nu}=l_\mu {\cal L}_\zeta \bar{F}^{\mu\nu} -{\cal J}^\nu~, \label{1.28}
	\\
	{\cal J}^\nu=-\frac 12 (i_1^\nu+fl^\nu l_\alpha {\bar j}^\alpha+\sigma_{i,i}l^\nu)~. \label{1.29}
\end{gather}
Also it follows from \eqref{1.24}, \eqref{1.26} that $\partial_v(\sigma_{i,i})=i_2^v$.  Thus the quantity ${\cal J}^\nu$ is solely determined by the current across the shockwave front. Like in the above case of scalar fields ${\cal J}^\nu$ indicates possible contributions to transition radiation when the source immediately contacts the shockwave.

We conclude, on quite general grounds, that normal components of the strength tensor
change  as
\begin{equation}\label{1.30}
l_\mu \hat{F}^{\mu\nu}\simeq \bar{\chi}(u) \left(l_\mu {\cal L}_\zeta \bar{F}^{\mu\nu}-{\cal J}^\nu\right)~.
\end{equation}
Outside domains, where sources contact the shockwave,
the change is generated by the Lie derivative along the supertranslation vector \eqref{1.14}.

The Cauchy data fix components $\hat{F}_{vi}$ and $\hat{F}_{vu}$. In fact, 
$\hat{F}_{uv}$
is not independent since it follows from the Maxwell equations that
\begin{equation}\label{1.32}
 \hat{F}_{uv}=-\frac12\int_{-\infty}^v \partial_i \hat{F}_{vi}~ dv'~.
\end{equation}
Also it follows from the Bianchi identities that
\begin{equation}\label{1.33}
	\begin{split}
		\hat{F}_{ui}&=\int_{-\infty}^v \left(\partial_u \hat{F}_{vi}+\partial_i \hat{F}_{uv}\right)dv'~,
		\\
		\hat{F}_{ij}&=\int_{-\infty}^v \left(\partial_i \hat{F}_{vj}-\partial_j \hat{F}_{vi} \right)dv'~.
	\end{split}
\end{equation}
It is assumed that $\bar{F}_{\mu\nu}$ vanish as $v\to-\infty$ and so does $\hat{F}_{\mu\nu}$.
Since the perturbations satisfy homogeneous Maxwell equations, we come to the characteristic problem for the two components
\begin{gather}
	\Box \hat{F}_{vi}=0~~,~~u>\delta~~, \label{1.35}
	\\
	\hat{F}_{vi}={\cal L}_\zeta \bar{F}_{vi}=\partial_v( f \bar{F}_{vi})~~,~~u=\delta~~.\label{1.36}
\end{gather}
The rest components are determined by \eqref{1.32}, \eqref{1.33}.  Equations (\ref{1.35}), (\ref{1.36}) prove  
(\ref{i.1})-(\ref{i.3}) for the case of EM fields.

\section{Applications of the results}
\label{sec:5}

Cauchy data (\ref{2.11}), \eqref{1.36} demonstrate our main statement about
geometrical form of the gravitational memory in field systems after gravitational shockwaves (\ref{1.1}), (\ref{1.1a}). 
Given the Cauchy data perturbations can be found by either numerical or analytical methods for shockwaves produced by different 
sources. 

The fact that scalar fields and normal components of the Maxwell strength tensor $F_{\mu\nu}$ change in the same way as trajectories of observers crossing the shockwave front does not imply that field systems behave smoothly across the fronts. One can check 
that observers, after the GSW, measure jumps of $\partial_u \phi$, $F_{ui}$ and of tangent components of the field stress-energy tensors.
Thus, in general, one has two effects: conversion of non-stationary perturbations into an outgoing
radiation and a plane shockwave of the field itself accompanying the initial gravitational wave. There may also appear  
a spherical shockwave when the gravitational wave
hits the source of the field \cite{Fursaev:2024czx}.

The effect of GSW on EM fields can be potentially observable for astrophysical objects, such as magnetars, which create strong magnetic fields. It has been suggested in \cite{Fursaev:2025did} that GSW from compact ultra-relativistic objects
which move near a magnetar generate short EM pulses. These pulses can be a novel engine mechanism for fast radio bursts. The fast radio bursts
are experimentally observed bright pulses of radio frequencies which are not yet fully explained, see~\cite{Petroff:2021wug}, \cite{Zhang:2020eou}. 

Analogous effects are produced by null cosmic strings moving near magnetars \cite{Fursaev:2023lxq}. Null cosmic strings are hypothetical objects,  they are cosmic strings which move with the speed of light and have a finite energy per unit length. 
Formally the metric of a null string belongs to shockwave class (\ref{1.1}), (\ref{1.1a}), so our analysis applies to this case as well.
A number of other physical effects caused by null strings on EM fields are considered in \cite{Fursaev:2022ayo}.
In particular, it is shown that parts of an incoming EM wave crossing the string horizon (null hypersurface $u=0$) from different sides
of the string are refracted with respect to each other and leave behind the string
a wedge-like region of interference. 

We have not discussed here perturbations of gravitational fields generated by plane-fronted GSW.
For the case of null cosmic strings these perturbations have been found and described in \cite{Fursaev:2023oep}.
The late-time asymptotics of the perturbations have a form of outgoing gravitational waves.  They have a number of interesting properties at the future null infinity.
By using arguments developed above it would be interesting to extend results of \cite{Fursaev:2023oep} to GSW.

\section{In the memory of A.A. Starobinsky}
\label{sec:6}

I would like to say several words about Alexei Aleksandrovich Starobinsky in Dubna, a small city on the banks of the Volga river. I hope the remarks below disclose additional features of A.A.Starobinsky's personality.

Dubna is known as a location of the Joint Institute for Nuclear Research, or JINR, the famous international scientific organization. A.A. Starobinsky maintained close ties with Dubna for many years. From 2019 to 2023 he was a member of staff at the Bogoliubov Laboratory of Theoretical Physics of JINR. 
When the project of Dubna International
Advanced School of Theoretical Physics (DIAS-TH) was launched at BLTP JINR in 2003, Alexey Aleksandrovich first became its lecturer and then the principal researcher.  Here interests and main activities of Alexei Aleksandrovich have been focused on the promotion of modern ideas in cosmology and astrophysics among students and young physicsts coming to Dubna  from all over the world. He was deeply involved in preparation of international schools and meetings, by working on their programmes and inviting best speakers.  His talks on current trends in cosmology always attracted many people from JINR. Thanks to his efforts and authority, these schools
were truly advanced.

As is known, JINR is also the host of the Russian mega-science project, a heavy ion collider called NICA.  When A.A.Starobinsky also became a Chair of the Russian Gravitational Society, he, Mikhail Vasil'evich Sazhin (1951-2023) and the author of this paper had organized the Workshop ``The Russian National Experiment in Field of Gravitational Physics''. The Worshop took place in May 2018 at the Dubna State University, also located in Dubna.  The purpose of the meeting was to discuss a Russian state-funded experimental programme in gravity, by analogy with mega-science projects in other areas of physics. Our initiate has been wellcome by the leading russian physicists and technical experts during this meeting. 

Alexey Aleksandrovich strongly supported the need of the national gravitational experimental programme.  I hope the work he started in Dubna may be revived in future.

\end{document}